\documentclass[twocolumn,pra,aps,showpacs,nofootinbib]{revtex4-1}
\usepackage{bm,multirow,amssymb,amsbsy,amsmath,graphicx,bbold,color}
\makeatletter
\usepackage{pifont}
\makeatother

\usepackage{color}
\newcommand{\ket}[1]{\vert #1 \rangle}
\newcommand{\bra}[1]{\langle #1 \vert}
\newcommand{\ES}{{\scriptscriptstyle E}}
\newcommand{\SP}{{\scriptscriptstyle S}}
\newcommand{\SE}{{\scriptscriptstyle S\!E}}
\newcommand{\HH}{{\hbox{\small HH}}}
\newcommand{\VV}{{\hbox{\small VV}}}
\begin{document}
\title{Quantum probes to assess correlations in a composite system}

\author{Andrea Smirne $^{1,2}$, Simone Cialdi $^{1,2}$, Giorgio Anelli $^1$, Matteo G.A. Paris $^{1,3}$, Bassano Vacchini $^{1,2}$}

\affiliation{ $^1$ \mbox{Dipartimento di Fisica, Universit{\`a} degli Studi di Milano, Via Celoria 16, I-20133 Milan, Italy}\\
$^2$ \mbox{INFN, Sezione di Milano, Via Celoria 16, I-20133 Milan, Italy}\\
$^3$ CNISM, Udr Milano, I-20133 Milan, Italy }


\begin{abstract}

We suggest and demonstrate experimentally a strategy to obtain 
relevant information about a composite system by only performing 
measurements on a small and easily accessible part of it, which we call quantum probe. 
We show in particular how
quantitative information about the angular correlations of couples of
entangled photons generated by spontaneous parametric down conversion is
accessed through the study of the trace distance between two polarization
states evolved from different initial conditions. After estimating the
optimal polarization states to be used as quantum probe, we provide a
detailed analysis of the connection between the increase of the trace distance
above its initial value and the amount of angular correlations.
\end{abstract}
\maketitle

The control of quantum systems plays a basic role in the experimental 
investigation of the predictions of quantum theory as well as 
in the development of quantum technologies for applications.
Indeed, great attention has been recently payed to engineering the dynamics 
of quantum systems in order to properly generate, manipulate and exploit significant quantum features 
 \cite{Myatt2000,Kraus2004,Chou2005,Verstraete2009,Barreiro2011,Krauter2011}.

Consider a large quantum system whose full characterization is only 
partially feasible or requires complex measurement schemes.
In such a case, it is crucial
to develop effective strategies in order to assess 
relevant pieces of information about the overall system 
by only monitoring a small part, which then acts as a probe.
A natural procedure is to control
the interaction of the small subsystem with the rest of the total system 
in such a way that the former can encode the information of interest.
Here, we provide an explicit example of this strategy in an all-optical
setup, where the system under study consists of entangled couples of
photons generated by spontaneous parametric down conversion (SPDC) \cite{Hong1985,Joobeur1994,Kwiat1999,Gerry2005}.
By properly engineering the interaction between polarization and 
momentum degrees of freedom of the photons via a 1D spatial light modulator (SLM),
we can access some information regarding the momentum correlations
between the two photons by simply performing
visibility measurements on the polarization degrees of freedom.

As specific figure of merit, we exploit the trace distance between polarization states. 
As we shall see, an increase of the trace distance above its initial value allows 
to detect some information on momentum correlations,
which has moved to the polarization degrees of freedom 
thanks to the engineered interaction. The
trace-distance analysis of quantum dynamics has been recently introduced, leading
to important results concerning the non-Markovianity of a quantum
dynamics
\cite{Breuer2009,Apollaro2011,Liu2011,Vacchini2011,Breuer2012,Smirne2013},
the characterization of the presence of initial correlations between the quantum system
and its environment
\cite{Laine2010b,Smirne2011,Dajka2011,Gessner2011}, the relevance of
non-local memory effects \cite{Laine2012,Liu2012,Laine2012b} and the reservoir engineering
in ultracold gases \cite{Haikka2011,Haikka2012,Haikka2013}.

The paper is structured as follows. In the next Section
we describe the physical system we are going to
investigate and present the details of the experimental
apparatus. In Section II we illustrate the trace-distance approach
to the dynamics of an open system and present the details of the
calculations of its evolution for different angular and polarization
states. Section III is devoted to illustrate the experimental results
about the optmization of the probe and the link between the behavior of trace distance 
and the initial correlations in the angular degrees of freedom.
Finally, Section IV closes the paper with some concluding remarks.

\section{The physical system and the experimental apparatus}\label{sec:exs}
Our overall system consists of couples of entangled photons generated by SPDC in a two-crystal geometry \cite{Kwiat1999}.
The couples are detected
along two beams, named signal and idler,
which are centered around the directions fixed by the phase matching condition. 
The two-photon state generated by SPDC can be written
\begin{eqnarray}
\label{eq:totall}
 |\psi\rangle = && \int d \omega_p d \omega_s d \theta_s d \theta_i A(\omega_p) 
\tilde{F}(\Delta k_{\perp}) {\mbox{Sinc}}( \Delta k_{\parallel} L/2) \nonumber\\
&&\times\left[\cos{\alpha}\ket{{\small{H,\theta_s, \omega_s}}}\ket{{\small{H,\theta_i,\omega_p-\omega_s}}}\right.\nonumber\\
&& \left.+e^{i \Phi(\omega_p, \theta_s, \theta_i)}\sin{\alpha}
\ket{{\small{V,\theta_s, \omega_s}}}\ket{{\small{V,\theta_i,\omega_p-\omega_s}}}\right], 
\end{eqnarray}
where up to first order in frequency and angle, 
\begin{eqnarray}
\Delta k_{\parallel} &=& -\frac{\omega_p^0 \theta^0}{2 c}(\theta_s+\theta_i)\nonumber\\
\Delta k_{\perp} &=& \frac{\omega_p^0}{2 c} (\theta_s-\theta_i) + \frac{2 \theta^0 \omega_s}{c}.\label{eq:kk}
\end{eqnarray}
Here, $\omega_p$ is the shift of the pump frequency with respect to the central frequency $\omega^0_p (405nm)$, 
$\theta_s$ and $\omega_s$ ($\theta_i$ and $\omega_i=\omega_p-\omega_s$) are the signal (idler)
angle and frequency shift with respect to the phase matching condition, $\theta_s^0=\theta_i^0\equiv\theta^0=3^{\circ}$ and $\omega^0_s=\omega^0_i= \omega^0_p/2$, 
while $\ket{{\small{P,\theta,\omega}}}$ denotes the single-photon state
with polarization ${\small{P = H,V}}$, angle $\theta$ and frequency $\omega$.
Moreover, $A(\omega_p)$ is the spectral amplitude of the pump, $\tilde{F}(\Delta k_{\perp})$ is
the Fourier transform of its spatial amplitude
and the ${\mbox{Sinc}}(\Delta k_{\parallel} L/2)$ function arises due to
the finite crystal size ($L=1mm$) along the longitudinal direction. 
The two-crystal geometry implies that the polarization degrees of freedom of the two photons are entangled and it further
introduces the phase term 
$\Phi(\omega_p, \theta_s, \theta_i)$,
which is due to the different
optical paths followed by the couples of photons generated in the first and in the second crystal
\cite{Rangarajan2009,Cialdi2010a,Cialdi2012}. To first order this term reads $\Phi(\omega_p, \theta_s, \theta_i) \approx \Phi_0 + \Delta \tau \omega_p + \kappa \theta_s +\eta \theta_i$.
Finally, the probabilities of generating $\ket{\VV}$ or $\ket{\HH}$
photons, $\sin^2{\alpha}$ and $\cos^2{\alpha}$ respectively, are determined by the polarization
of the incident laser. 

The overall state in Eq.(\ref{eq:totall}) fixes in particular
the correlations between signal and idler angular degrees of freedom. 
By properly engineering the two-photon evolution, relevant information about these angular correlations 
gets encoded into the polarization degrees of freedom and then can be easily accessed. 
In fact, through the SLM we can impose an arbitrary polarization- and
position-dependent phase shift to the two-photon state in Eq.(\ref{eq:totall}). 
On the one hand, a linear phase $\overline{\Phi} \equiv -\Phi_0 - \kappa \theta_s -\eta \theta_i$ 
is set to
offset the corresponding terms in the first-order expansion of $\Phi(\omega_p, \theta_s, \theta_i)$ \cite{Rangarajan2009,Cialdi2010a,Cialdi2012,Cialdi2010b}.
On the other hand, a further linear phase on both signal and idler beams emulates a time
evolution of the two-photon state \cite{Cialdi2011}. 

\begin{figure}[ht!]
\includegraphics[width=1.05\columnwidth]{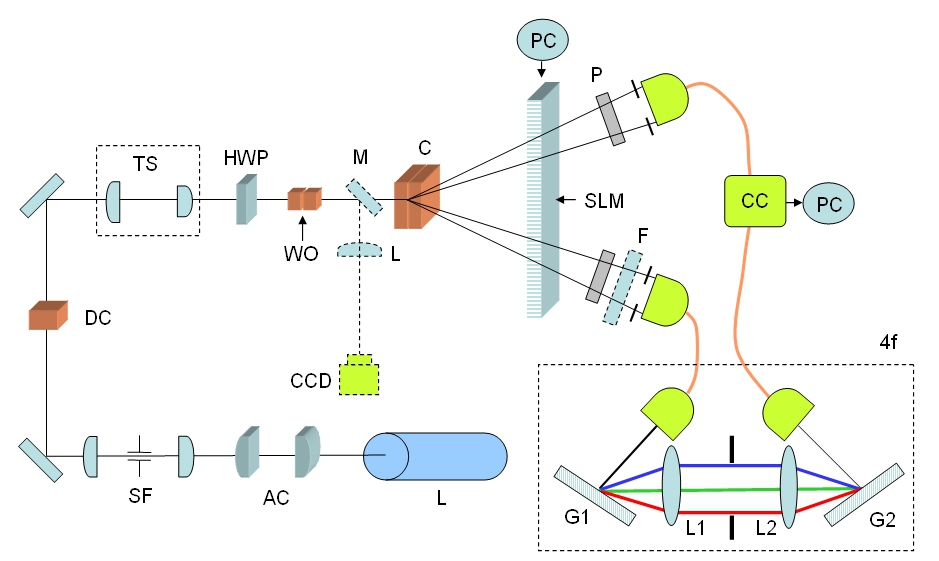}
\caption{(Color online) Schematic diagram of the experimental
setup, as described in Sect.\ref{sec:exs}. }\label{fig:1}
\end{figure}

The experimental setup is shown in Fig. \ref{fig:1}. A linearly
polarized cw 405 nm diode laser (Newport LQC405-40P) passes through two cylindrical lenses, which compensate
beam astigmatism (AC), then through a spatial filter (SF) composed
by two lenses and a pin-hole in the Fourier plane to obtain
a Gaussian profile by removing the multimode spatial
structure of the laser pump. Finally, a telescopic
system (TS) prepares a beam with the proper radius and
divergence. A couple of 1 mm beta-barium borate crystals (C),
cut for type-I downconversion, with optical axis
aligned in perpendicular planes, are used as a source
of polarization and momentum entangled photon pairs
with $\theta^0=3^{\circ}$.
We use a compensation crystal on the pump (DC) \cite{Cialdi2008},
which acts on the delay time between the vertical and
horizontal polarization, and a couple of thin crystals ($0.5 mm$) for the
spatial walk-off compensation (WO).
An interference filter or a long pass filter (F) is put on the signal path to select the spectral
width of the radiation (10nm or 45nm).
In order to obtain different spectral widths or a particular spectral profile, we 
use a 4f optical system after the coupler on the signal path.
The 4f system consists of two gratings (G1 and G2) of $1200 lines/mm$ and two achromatic lenses (L1 and L2)
with $f=35mm$.
The distance between the lenses and the grating is $f$ and the distance between the two lenses is 2f. In this
configuration, in between the two lenses the spectral components are focalized and well separated, 
so that it is possible to put here a slit to select the wanted spectral width.
An SLM, which is a liquid
crystal phase mask $(64 \times 10 mm)$ divided in 640 horizontal
pixels each $d=100 \mu m$ wide, is set before the
detectors, at 310 mm from the generating crystals, in order to introduce the spatial phase function. 
When the mirror (M) is switched on the radiation path a cylindrical lens  (L)
generates the Fourier Transform profile of the pump at his focal distance (1m), where a CCD camera is located.
A couple of polarizer (P) is used to measure the visibility of the entangled state.

\section{Trace-distance analysis} 
\subsection{General strategy}
The trace distance between two quantum states $\rho^1$ and $\rho^2$ is defined as 
\begin{equation}\label{eq:tdd}
D(\rho^1, \rho^2) = \frac{1}{2} \mbox{Tr}\left|\rho^1-\rho^2\right|=\frac{1}{2}\sum_k|x_k|,
\end{equation} 
with $x_k$ eigenvalues of the traceless operator
$\rho^1-\rho^2$, and it is a metric on the space of physical states such
that it holds $0\leq
D(\rho^1, \rho^2) \leq1$.
The physical meaning of the trace distance lies in the fact that it measures
the distinguishability between two quantum states \cite{Fuchs1999}. 
As a consequence, given an open quantum system $S$ interacting with an environment $E$ \cite{Breuer2002},
any variation of the trace distance of two open system's states  $D(\rho^1_\SP(t), \rho^2_\SP(t))$
can be read in terms of an exchange of information between the open system and the environment \cite{Breuer2009,Laine2010b,Breuer2012}.
Here, $\rho^1_\SP(t)$ and $\rho^2_\SP(t)$ are open system's states evolved from different
initial total states $\rho^1_\SE(0)$ and $\rho^2_\SE(0)$ through the relation
$\rho^k_\SP(t)=\mbox{tr}_\ES\left\{U(t) \rho^k_\SE(0) U^{\dag}(t)\right\}$, $k=1,2$,
where the total system $SE$ is assumed to be closed and hence evolves
through a unitary dynamics $U(t)$  \cite{Breuer2002}. In particular, if there are no initial system-environment
correlations,
$
\rho^1_\SE(0)=\rho^1_\SP(0)\otimes\rho^1_\ES(0)$ and $\rho^2_\SE(0)=\rho^2_\SP(0)\otimes\rho^2_\ES(0)$, 
an increase of the trace distance above its initial value,
\begin{equation}\label{eq:ris}
 D(\rho^1_\SP(t), \rho^2_\SP(t)) > D(\rho^1_\SP(0), \rho^2_\SP(0)),
\end{equation}
witnesses the difference of the two initial environmental states, i.e.
$\rho^1_\ES(0) \neq \rho^2_\ES(0)$ \cite{Laine2010b,Smirne2010}.
This relation already shows how the trace distance between open system's states allows
to access nontrivial information regarding the environment. 
More specifically, in the following we present a quantitative link
between the trace distance behavior and the environmental correlations.

\begin{figure}[ht!]
\includegraphics[width=.97\columnwidth]{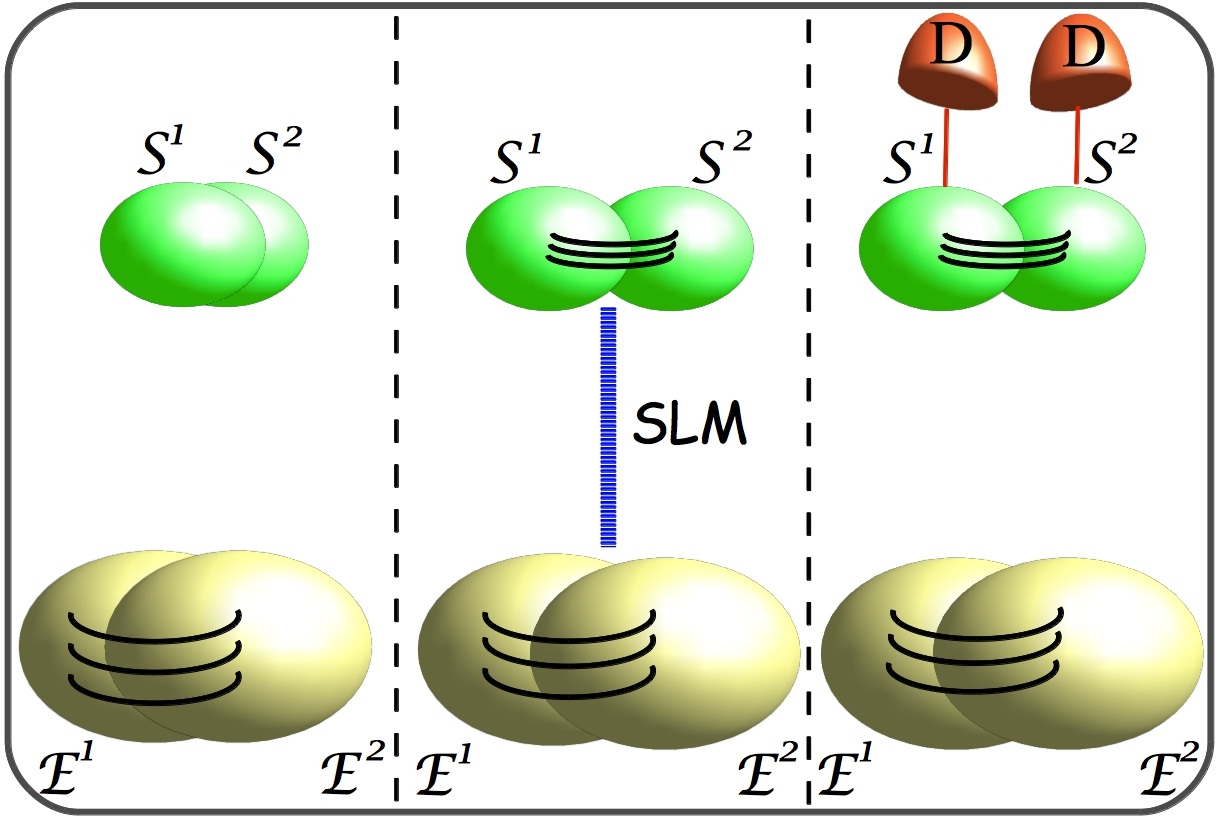}
\caption{(Color online) 
Logical scheme of the experiment. In the first stage system and
environment are uncorrelated, and the environmental states ($E^1$ and $E^2$)
differ due to correlations.
The wiring represents the information about these different correlations.
In the second stage system and environment are
coupled through the SLM, so that information on the environmental correlations
is transferred to
the couple of system states ($S^1$ and $S^2$), making them more distinguishable,
and it is finally read out through the detector (D) acting on the system only, in the third and
final stage. Note that the two states refer to
two distinct runs of the experiment.}\label{fig:2}
\end{figure}

In view of the trace-distance analysis,
our physical system can be characterized as follows.
The polarization degrees of freedom are the open system $S$
and the angular degrees of freedom the corresponding environment $E$. 
The latter are
in turn manipulated by varying the divergence of the pump, as well as
by selecting the frequency-spectrum width of the two-photon state generated by SPDC.
We therefore study the evolution of the trace distance
between two polarization states evolved
from different initial $SE$ states, which can be considered product states
thanks to the compensation of the phase term introduced by the SLM.  In particular, we investigate how
the trace-distance evolution of the polarization degrees of freedom, which are a
small and easily accessible component of the total system, is sensitive to the different angular correlations
within $\rho^1_\ES$ and $\rho^2_\ES$, thus allowing
to assess this characteristic feature of the overall two-photon state.
A logical scheme of the experiment is depicted in Fig.\ref{fig:2}.

Let us emphasize that our apparatus exploits
all the degrees of freedom of the photons generated by SPDC: 
polarization degrees of freedom as the open system, 
angles as the environment and frequencies, together with the spatial properties of the pump, as a tool to vary
the correlations within the environment.

\subsection{Trace distance evolution for different angular and polarization states}
In our apparatus, the angular state after the compensation of the phase
through the SLM can be described, setting $\bm{\theta}\equiv (\theta_s,\theta_i)$, as
\begin{equation}
  \label{eq:rhoe}
  \rho_\ES = \int d\bm{\theta}d\bm{\theta}' h(\bm{\theta};\bm{\theta}')
\ket{\small{\bm{\theta}}}\bra{\small{\bm{\theta}'}},
\end{equation}
with 
\begin{equation}\label{eq:rhoee}
h(\bm{\theta}; \bm{\theta}') \equiv  {\mbox{Sinc}}(\bm{\theta}){\mbox{Sinc}}(\bm{\theta}') \int_{\Omega_s} d \omega_s 
\tilde{F}(\Delta k_{\perp})\tilde{F}^*(\Delta k_{\perp}'). 
\end{equation}
The influence of the pump spectrum on the angular state can be neglected \cite{Cialdi2010a,Cialdi2010b}
and the integration over $\omega_s$ is performed on the frequency interval $\Omega_s$
selected  by the filter or the $4f$ setup on the signal path.
The joint probability distribution $P(\bm{\theta})\equiv
h(\bm{\theta}; \bm{\theta})$
determines the angular correlations, which can be quantified as  
\begin{equation}\label{eq:corr}
 C=\frac{\langle \theta_s \theta_i\rangle-\langle \theta_s\rangle\langle\theta_i\rangle}{\sqrt{V_{s} V_i}},
\end{equation}
with $V_{j}=\langle \theta^2_j\rangle-\langle \theta_j\rangle^2$ variance of the 
angular distribution $P(\theta_j)$, $j=s,i$.
In particular, given a collimated beam with a large pump waist,
so that the transverse momentum is nearly conserved,
the signal and idler angles are the more correlated
the less the selected frequency spectrum is wide.
On a similar footing, for a $10 nm$-spectrum and a fixed pump
waist, the correlation of the angular degrees of freedom
grows with decreasing pump divergence.
Thus, we can control the initial correlations of the environment by selecting
the frequency spectrum
of the two-photon state or the divergence of the
pump. 

The polarization state after the purification trough the SLM reads
\begin{equation}\label{eq:rhos}
 \rho_\SP = \gamma \ket{\psi}\bra{\psi} + (1-\gamma)\rho^m,
\end{equation}
where 
\begin{equation}\label{eq:rhoss}
 \ket{\psi}=\cos{\alpha}\ket{\HH}+\sin{\alpha}\ket{\VV},
\end{equation}
see Eq.(\ref{eq:totall}), 
is a pure state and $\rho^m = \cos^2{\alpha}\ket{\HH}\bra{\HH}+\sin^2{\alpha}\ket{\VV}\bra{\VV}$ the corresponding mixture.
In our setting, we can control $\alpha$ 
via the polarization of the pump,
while $\gamma$ can be modified  by changing
the crystal along the pump which precompensates the delay time
due to the two-crystal geometry \cite{Cialdi2008}.
The purity $p=\mbox{Tr}\rho_\SP^2$ of $\rho_\SP$ is
\begin{equation}\label{eq:pur}
p=1-\frac 12  (1-\gamma^2) \sin^2(2\alpha), 
\end{equation}
 whereas its concurrence $\mathcal{C}$ \cite{Wootters1997} reads
\begin{equation}\label{eq:conc}
 \mathcal{C}=\gamma|\sin(2\alpha)|.
\end{equation}

In the following, we will compare the evolution of polarization states
evolved from different initial states $\rho^k_\SP(0)\otimes\rho^k_\ES(0)$, $k=1,2$.
The two initial open system's states $\rho^k_\SP(0)$ have polarization parameters $\alpha_k$ and $\gamma_k$, 
see Eqs.(\ref{eq:rhos}) and (\ref{eq:rhoss}),
while the two initial environmental states $\rho^k_\ES(0)$ 
have angular amplitudes $h_k(\bm{\theta}; \bm{\theta}')$, see Eqs.(\ref{eq:rhoe}) and (\ref{eq:rhoee}),
and then joint angular probability distributions $P_k(\bm{\theta})=h_k(\bm{\theta}; \bm{\theta})$.
In particular, we impose through the SLM 
a linear phase which can be described 
through the unitary operators 
\begin{equation}\label{eq:U}
 U(\beta)\ket{{\small{V \theta_s}}}\ket{{\small{V \theta_i}}} = e^{i \beta (\theta_s-\theta_i)} 
\ket{{\small{V \theta_s}}}\ket{{\small{V \theta_i}}},
\end{equation}
where $\beta$  is the evolution parameter.
The polarization states for a generic value
of $\beta$ are  then
\begin{equation}\label{eq:rhoskb}
 \rho^k_\SP(\beta) =  \frac{\epsilon_k(\beta)}{\sin(2\alpha_k)} \ket{\psi_k}\bra{\psi_k} + \left(1- \frac{\epsilon_k(\beta)}{\sin(2\alpha_k)}\right)\rho^m_k,
\end{equation}
where $\ket{\psi_k}=\cos{\alpha_k}\ket{\HH}+\sin{\alpha_k}\ket{\VV}$ and
\begin{equation}\label{eq:epsk}
 \epsilon_k(\beta) =\gamma_k \sin(2\alpha_k) \int d \theta_s d \theta_i 
e^{i \beta (\theta_s-\theta_i)} P_k(\theta_s,\theta_i),
\end{equation}
which is a real function of $\beta$ since the joint probability distribution is symmetric
under the exchange $\theta_s \leftrightarrow \theta_i$.
It is worth emphasizing that the absolute value of $\epsilon_k(\beta)$ equals
the concurrence as well as the interferometric visibility of the state $\rho^k_\SP(\beta)$. 
In particular, we measure the visibility by counting the coincidences with polarizers set at
$45^{\circ},45^{\circ}$ and at $45^{\circ},-45^{\circ}$, see
\cite{Cialdi2009} for further details.
Moreover, by virtue of the specific evolution obtained through the SLM, 
$\epsilon_k(\beta)$ is fixed by the Fourier transform of the spatial
profile $|\tilde{F}(\Delta k_{\perp})|^2$, which at first order is a
function of $\theta_s-\theta_i$  and $\omega_s$, see Eq.(\ref{eq:kk}), thus depending
on both the pump divergence and the selected frequency spectrum.
Thus the engineered evolution, see Eq.(\ref{eq:U}), guarantees that
the interferometric visibility is sensitive to the different angular correlations
in the environment.

Finally, the trace distance
$D(\beta)\equiv D(\rho^1_\SP(\beta), \rho^2_\SP(\beta))$
between the polarization states
is simply given by, see Eqs. (\ref{eq:tdd}) and (\ref{eq:rhoskb}),
\begin{equation}\label{eq:dbeta}
D(\beta) =\sqrt{(\cos^2\alpha_1-\cos^2\alpha_2)^2+\left(\epsilon_1(\beta)-\epsilon_2(\beta)\right)^2/4}. 
\end{equation}

\section{Experimental results}
\subsection{Characterization of the probe} 
\begin{figure}[ht!]
\includegraphics[width=0.99\columnwidth]{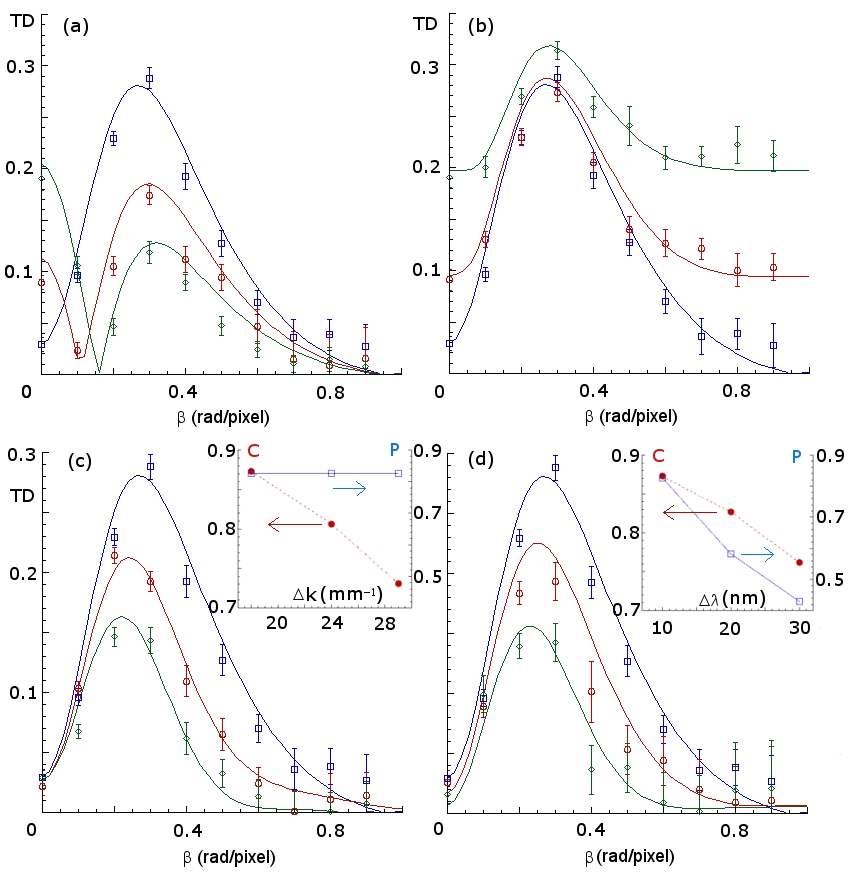}
\caption{(Color online) Trace distance $D(\beta)$ versus the
evolution parameter $\beta$ for different couples of initial system states
and different environmental correlations, see Eq.(\ref{eq:corr}), showing
how the trace-distance growth is sensitive to the environmental correlations. (a) and (b):
$\rho^1_\ES(0)$ and $\rho^2_\ES(0)$ are kept fixed ($\Delta k_1=\Delta k_2=18 mm^{-1}$, 
$\Delta \lambda_1 = 45 nm$ and $\Delta \lambda_2= 10 nm$),
$\rho^1_\SP(0)$ is fixed with $\alpha_1=\pi/4$ and $\gamma_1=0.91$
(mainly due to contributions to the phase in Eq.(\ref{eq:totall}) which are not compensated to the first order),
$\rho^2_\SP(0)$ corresponds to $\alpha_2=\pi/4$ and $\gamma_2 = 0.96$ (blue line), 
$0.73$ (red line), $0.52$ (green line) in (a), 
while $\gamma_2=0.96$ and
$\alpha_2 = \pi/4$ (blue line), $0.675$ (red line), $0.575$ (green line) in (b). (c) and (d): $\rho^1_\SP(0)$ and $\rho^2_\SP(0)$ are fixed
($\alpha_1=\alpha_2=\pi/4$, $\gamma_1 = 0.91$ and $\gamma_2=0.96$),
$\rho^1_\ES(0)$ is fixed with $\Delta \lambda_1 = 45 nm$ and $\Delta k_1 =18 mm^{-1}$, while $\rho^2_\ES(0)$
corresponds to $\Delta \lambda_2 = 10 nm$ and $\Delta k_2= 18 mm^{-1}$ (blue line), $24 mm^{-1}$ (red line), $29 mm^{-1}$ (green line) in (c) 
and to $\Delta k_2 =18 mm^{-1}$ and $\Delta \lambda_2 = 10 nm$ (blue line), $20 nm$ (red line), $30 nm$ (green line) in (d).
The insets in (c) and (d) show the angular correlations (red line) and the purity (blue line) of $\rho^2_\ES(0)$ 
as a function of $\Delta k_2$ (in (c)) or $\Delta \lambda_2$ (in (d)).
Experimental data are reported with their error bars,
the solid lines represent the theoretical predictions.}\label{fig:3}
\end{figure}

As a first step, we show how the choice of the initial 
polarization states, $\rho^1_\SP(0)$ and $\rho^2_\SP(0)$,
influences in a critical way whether the subsequent trace-distance evolution
is an effective probe of the different angular correlations in the two initial angular states.
To this aim, 
we fix $\rho^1_\ES(0)$ and $\rho^2_\ES(0)$ as 
the states corresponding to $\Delta \lambda_1 = 45 nm$ and $\Delta \lambda_2 = 10 nm$, respectively, that
is a weakly and a strongly correlated angular state, while
we consider different couples of initial polarization states.
To this aim we set  $\alpha=\pi/4$ for both $\rho^1_\SP(0)$ and $\rho^2_\SP(0)$, and keep $\gamma_1$ fixed, while we vary $\gamma_2$
by inserting different precompensation crystals.
Specifically, we exploit a $3 mm$ crystal to fully compensate
the delay time $\Delta \tau$ \cite{Cialdi2008}, a $1 mm$ crystal to partially 
compensate it and we also consider the case without any precompensation crystal.
The experimental data, together with the theoretical prediction obtained by Eq.(\ref{eq:dbeta}), 
are shown in Fig.\ref{fig:3}.(a). For high values of $\gamma_2$,
the trace distance between polarization states actually satisfies Eq.(\ref{eq:ris})
and then witnesses the different initial conditions in the angular degrees of freedom. 
The information due to the differences
in $\rho^1_\ES(0)$ and $\rho^2_\ES(0)$ flows to the polarization
degrees of freedom because of the engineered interaction. Thus,
one can access through simple visibility measurements on the open
system some information which was initially outside it.
On the other hand, the revival of the trace distance above its initial value decreases
with the decreasing of $\gamma_2$, and
for low enough values of $\gamma_2$ the trace distance remains below its initial value
for the whole evolution. 
The loss of purity and entanglement due to a decrease of the parameter $\gamma$ 
in the initial polarization states can prevent the subsequent trace distance
from being an effective probe of the different correlations in the angular states.  

The relative weight of vertically and horizontally polarized
photons generated by SPDC is determined by the parameter $\alpha$, which can be controlled
by properly rotating a half-wave plate set on the pump beam.
In Fig.\ref{fig:3}.(b) we report the experimental data and theoretical
predictions of the trace-distance behavior for a given value of
$\alpha_1$ as well as fixed $\gamma_1$ and $\gamma_2$, while considering different values of $\alpha_2$.
One can see that, even if the growth of the trace distance
above its initial value decreases with the decreasing of $\alpha_2$, 
it is still visible also for a sensible
imbalance between vertically and horizontally polarized photons.
Indeed, for a fixed $\gamma<1$ the decrease of $\sin(2\alpha)$
in the polarization states $\rho^k_\SP(0)$
corresponds to a decrease of the concurrence, but to an increase of the purity, see Eqs.(\ref{eq:pur}) and (\ref{eq:conc}).
Contrary to what happens for a decrease of the parameter $\gamma_2$, see Fig.\ref{fig:3}.(a),
the open system always recovers the information initially outside it
from the very beginning of its evolution and the trace-distance maximum
increases with the increasing of the initial distinguishability
between the two polarization states.

\subsection{Trace distance as a witness of initial correlations in the angular degrees of freedom}
The analysis of the previous paragraph
shows that the optimal probe of the angular correlations
is achieved by exploiting the highest amount of
purity and entanglement of the polarization degrees of freedom available
within our setting. Now, we study
how this optimal probe reveals changes in the angular correlations.
Hence, we fix $\rho^1_\SP(0)$, $\rho^2_\SP(0)$
and we investigate the trace-distance evolution $D(\rho^1_\SP(\beta),\rho^2_\SP(\beta))$
for different couples of initial angular states.
In particular, we take as reference environmental state
$\rho^1_\ES(0)$ the state
with weak angular correlations, which is obtained by means of a 
collimated beam and a $45nm$-spectrum. We compare the evolution
of the subsequent polarization state $\rho^1_{\SP}(\beta)$ with
the evolution of a state $\rho^2_\SP(\beta)$ evolved in the presence
of strong initial angular correlations in $\rho^2_\ES(0)$. We repeat this procedure
by changing the amount of correlations $C$ in $\rho^2_\ES(0)$, see Eq.(\ref{eq:corr}), thus studying
the connection between $C$ and the effectiveness of the quantum
probe of the angular correlations quantified by the increase of the trace distance
above its initial value. 
 
\begin{figure}[ht!]
\includegraphics[width=.9\columnwidth]{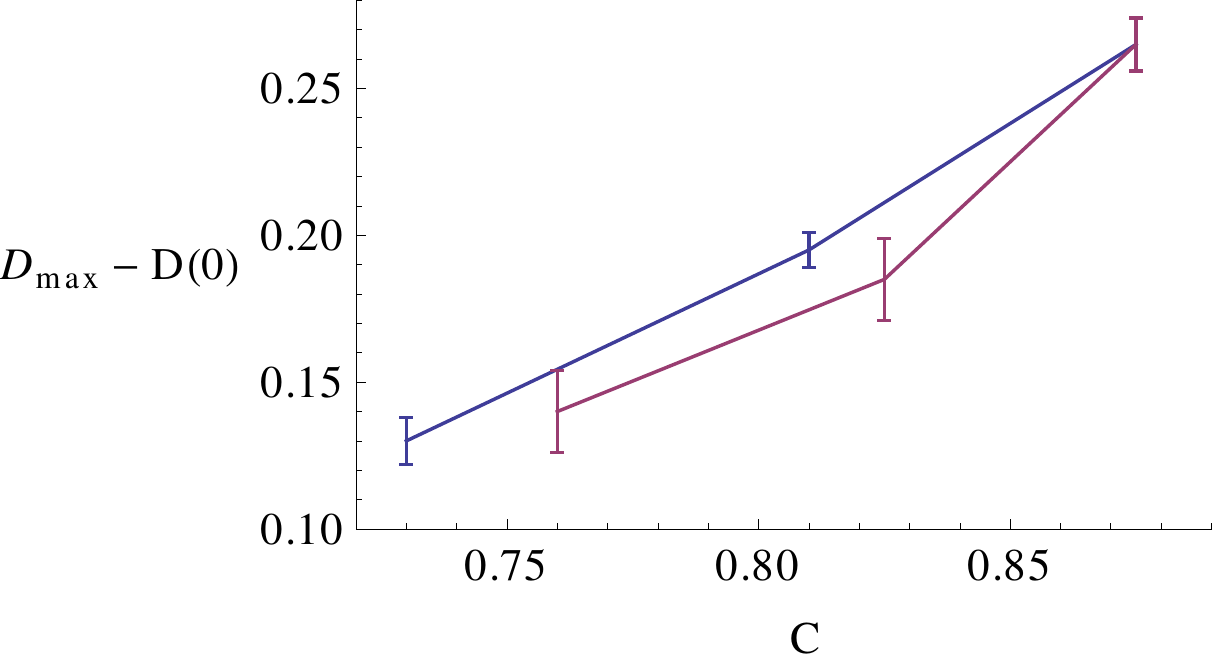}
\caption{(Color online) Experimental values of the maximum increase of the trace distance
above its initial value as a function of the angular correlations $C$, see Eq.(\ref{eq:corr}).
The experimental data are referred to the behavior of the trace distance
for different beam divergences (blue line, compare with Fig.\ref{fig:3}.(c)) or
different widths of the frequency spectrum (red line, compare with Fig.\ref{fig:3}.(d)).}\label{fig:4}
\end{figure}

In Fig.\ref{fig:3}.(c), one can see the experimental data and theoretical prediction
concerning the different trace-distance evolutions $D(\rho^1_\SP(\beta),\rho^2_\SP(\beta))$
which correspond to the different beam divergences exploited, together with a $10nm$-spectrum, 
in the preparation of the environmental state $\rho^2_{\ES}(0)$.
The divergence is enlarged by suitably
setting a telescopic system of lenses, so that the FWHM $\Delta k$ of $|\tilde{F}(\Delta k_{\perp})|^2$ is increased, 
while the $220 \mu m$ spot on the generating crystals is kept fixed \cite{Cialdi2012}.
The increase of the trace distance above its initial value
grows with the angular correlations $C$ in $\rho^2_{\ES}(0)$.
The behavior of $D(\rho^1_\SP(\beta),\rho^2_\SP(\beta))$ 
indicates that, for the specific choice of $\rho^1_\SP(0)$ and $\rho^2_\SP(0)$,
the trace distance can actually witness even a small difference in the angular correlations of
$\rho^1_\ES(0)$ and $\rho^2_\ES(0)$. 
The direct connection between the trace distance and
the correlations in the environment is further shown in Fig.\ref{fig:4}, where the difference between the maximum
and the initial
value of the trace distance is plotted as a function of the angular correlations.
The experimental data point out that the probe represented by the trace distance
is sensitive to the different amount of correlations within the environment,
which is indeed not a priori entailed 
by Eq.(\ref{eq:ris}).

As a further check of the connection between angular correlations
and the increase of the trace distance above its initial value,
we take into account environmental states $\rho^2_\ES(0)$
in which the angular correlations are modified
by selecting different frequency spectra of the two-photon state. Besides affecting the angular
correlations, this also influences the purity of $\rho^2_\ES(0)$.
As it may be inferred from Eqs.(\ref{eq:rhoe}) and (\ref{eq:rhoee}), a wider frequency spectrum
implies a lower angular purity, as a consequence of the fact
that the pure state generated by SPDC in Eq.(\ref{eq:totall})
also involves the frequencies. However,
the trace-distance evolution does not keep track of the purity
of the environmental state. In particular, the growth of the trace distance above its initial value
is not affected by the different purities of $\rho^2_\ES(0)$
in the two situations, but is determined by the amount of angular correlations $C$,
see the insets in Fig.\ref{fig:3}.(c), (d) and Fig.\ref{fig:4}.
Indeed, this can be explained through Eqs.(\ref{eq:epsk}) and (\ref{eq:dbeta}):
the trace distance solely depends
on the angular probability distribution $P(\theta_s,\theta_i)$,
while it is independent of angular coherences.

\section{Conclusion}
We have theoretically described and experimentally demonstrated a strategy
to assess relevant information about a composite system by only
observing a small and easily accessible part of it. By exploiting
couples of entangled photons generated by SPDC and engineering a
proper interaction by means of a SLM, we could reveal correlations
within the angular degrees of freedom of the photons by monitoring the
trace distance evolution between couples of polarization states.
After estimating the optimal probe, we have shown that the increase of
the trace distance between system states above its initial value
provides a signature of the amount of angular correlations in the
environmental states.

\acknowledgments
This work has been supported by the COST Action MP1006 and the MIUR Project FIRB LiCHIS-RBFR10YQ3H.

\end{document}